\documentclass[reprint,
showpacs,preprintnumbers,showkeys,nofootinbib, aps,prd]{revtex4-1}

\usepackage{amsmath}
\usepackage{amssymb}
\usepackage{amsfonts}
\usepackage{latexsym}
\usepackage{graphicx}
\usepackage{color}
\usepackage{soul}

\newcommand{\eq}[1]{(\ref{#1})}

\newcommand{\n}[1]{\label{#1}}

\def\beq{\begin{eqnarray}}
\def\eeq{\end{eqnarray}}
\def\ln{\,\mbox{ln}\,}

\def\Det{\,\mbox{Det}\,}

\def\tr{\,\mbox{tr}\,}

\def\Tr{\,\mbox{Tr}\,}

\def\ln{\,\mbox{ln}\,}                  
\def\tr{\,\mbox{tr}\,}                  
\def\Tr{\,\mbox{Tr}\,}                  
\def\Box{\square}                       
\def\Det{\,\mbox{Det}\,}                

\def\na{\nabla}                         
\def\pa{\partial}                       

\def\={\ =\ }

\def\al{\alpha}
\def\be{\beta}

\def\ga{\gamma}
\def\de{\delta}
\def\ep{\epsilon}

\def\ze{\zeta}

\def\la{\lambda}

\def\si{\sigma}
\def\om{\omega}
\def\ph{\varphi}

\def\Ga{\Gamma}
\def\De{\Delta}
\def\La{\Lambda}

\begin{document}

\preprint{PrePrint}

\title{Non-local form factors for curved-space antisymmetric fields}

\author{Tib\'erio de Paula Netto}
\email{tiberiop@fisica.ufjf.br}
\affiliation{Departamento de F\'{\i}sica, ICE,
Universidade Federal de Juiz de Fora, 36036-330, MG, Brazil}

\author{Ilya L. Shapiro}
\email{shapiro@fisica.ufjf.br}
\affiliation{Departamento de F\'{\i}sica, ICE,
Universidade Federal de Juiz de Fora, 36036-330, MG, Brazil}
\affiliation{Tomsk State
Pedagogical University and Tomsk State
University, Tomsk, 634041, Russia}

\date{\today}

\begin{abstract}
\noindent
In the recent paper Buchbinder, Kirillova and Pletnev presented
formal arguments concerning quantum equivalence of free
massive antisymmetric tensor fields of second and third rank to
the free Proca theory and massive scalar field with minimal
coupling to gravity, respectively. We confirm this result using
explicit covariant calculations of non-local form factors based
on the heart-kernel technique, and discuss the discontinuity of
quantum contributions in the massless limit.
\end{abstract}



\maketitle


\section{Introduction}
\label{Sect1}

The antisymmetric fields in four dimensions are interesting from
various  viewpoints. The most attractive part is that they emerge
naturally as effective fields after compactification of the
(super)string effective action \cite{1}-\cite{5}. Therefore,
the detection of these fields or their low-energy quantum effects
may be regarded as indirect detection of (super)string theory.
Naturally, the standard situation is that such fields emerge as
parts of the corresponding supermultiplets. At the same time, at
low energies the supersymmetry is supposed to be broken and
the mainstream approach is the soft symmetry breaking related
to the introduction of masses. Therefore, one can expect that the
antisymmetric fields can be massive. At the same time, due to
compactification of extra dimensions such a mass can be quite
small, and hence it is interesting to see what happens in the
massless limit, especially at the quantum level. Let us note that
the antisymmetric tensor fields have also interesting
applications to the construction of non-minimal Grand
Unification models \cite{Adler}, where the interface between
massless and broken-symmetry massive versions is one of the
main issues.

The quantum aspects of massless antisymmetric fields has
been explored in Refs.~\cite{8}-\cite{13} (also, both
massless and massive cases were explored within the
worldline approach \cite{Bastianelli-massless,Bastianelli-massive}).
In particular, there
was found a quantum equivalence with vector and scalar
fields (classical equivalence was established before in \cite{6}),
and was shown that the massless third-rank field has no physical
degrees of freedom \cite{7}, \cite{13} - \cite{16}. Indeed, the first 
work where the equivalence between Proca model and antisymmetric
second-rank field can be seen was published long before 
in 1960 \cite{Kemmer}. Taking these
results into account, one of the interesting questions is about the
possible discontinuity of the quantum effects of antisymmetric
fields in the massless limit. Recently, quantum theory of massive
antisymmetric fields was considered in Ref.~\cite{Buch-Kiri-Plet}.
In particular, it was shown that the mentioned models are equivalent
to vector and scalar massive fields, correspondingly. The equivalence
holds in curved space-time, and not only at the classical level, but
also in a semiclassical theory, that means the contributions of the
corresponding fields to the effective action of vacuum are identical
to the ones of vector and minimal scalar theories. The proof of
\cite{Buch-Kiri-Plet} is very general and is based on
$\ze$-regularization. However, this type of proofs is
always interesting to check by direct calculation, similar to
what was done for the Proca model \cite{BuGui}. In this
case one can detect the discontinuity of the massless limit
not only in the local divergent terms, but also in the
complicated non-local  contributions, which are typical for
the massive field. Another interesting aspect it that the proofs
of equivalence involve operations which are potentially
dangerous with respect to the non-local multiplicative
anomaly, which was previously detected only for fermionic
determinants \cite{QED-form}. It looks reasonable to to see
whether a similar situation takes place in the case of antisymmetric
fields and their quantum equivalence with massive vectors and
minimal massive scalars.

In the present work we will derive the one-loop form
factors using the heat-kernel technique based on the exact solution
by Barvinsky and Vilkovisky and Avramidi \cite{Bar-Vil90,Avramidi}.
Such a calculation was previously performed for various models,
including scalar field \cite{apco}, fermions and massive vectors
\cite{fervi}. The equivalence with the derivation by means
of Feynman diagrams has been shown in \cite{apco} and
more recently in \cite{CadeOmar}. Indeed, the heat-kernel
based method is much more economic and since the application
to antisymmetric fields is technically complicated, we chose
this approach.

The paper is organized as follows. In Sect.~\ref{o} we briefly review
the heat-kernel derivation of form factors according to \cite{apco}
and present the final results for massive scalars and vectors. The
second-rank antisymmetric tensor theory is worked out in
Sect.~\ref{sec-rank2}, and Sect.~\ref{sec-rank3} deals with the
third-rank case. Most of these two sections repeats the contents
of \cite{Buch-Kiri-Plet} and other references. The reason to include
this  is the intention to make the work self-consistent. Therefore,
this part is made as brief as possible, but at the same time we give
sufficient details. Finally, the Conclusions are drawn in
Sect.~\ref{Conc}.

\section{Derivation of the one-loop form factors}
\label{o}

Let us review the derivation of the one-loop effective action up to
the terms of the second order in curvatures. More details can be
found in Refs.~\cite{apco,fervi}.
The one-loop Euclidian effective action is given by the formula
\beq
\n{EA}
\bar{\Ga}^{(1)} =   \frac{1}{2} \Tr \ln \hat{H}
\,,
\eeq
where we assume the minimal form of the bilinear form of the action
\beq
\n{Bilinear}
\hat{H}
&=&
\hat{1}\Box  - \hat{1}m^2  + \hat{P} - \frac{\hat{1}}{6}\, R \,,
\eeq
where $\hat{1}$ is an identity matrix in the space of the fields of
interest and  $\hat{P}$ operator depends on the metric and possibly
other background fields. The commutator of the two covariant
derivatives acting on the corresponding fields are
$\,\hat{S}_{\mu\nu} = [\na_\mu,\na_\nu]$.

The effective action \eq{EA} can be presented as an integral in
the proper time $s$, involving the heat kernel $K(s)$,
\beq
\n{EA2}
\bar{\Ga}^{(1)} = - \frac{1}{2} \, \int_0^\infty \frac{ds}{s}
\, \Tr K(s)\,.
\eeq
The heat kernel can be expanded up to the second order in
the curvatures, namely Ricci tensor $R_{\mu\nu}$ and
scalar $R$, $\,\hat{S}_{\mu\nu}\,$ and $\,\hat{P}$. The
second-order solution has the form \cite{Bar-Vil90}
\beq
\n{Heat-Kernel}
\Tr K(s) &=& \frac{\mu^{2(2-\om)}}{(4 \pi s)^\om}
\int d^{2\om} x \, \sqrt{g} \, e^{-s m^2}
\tr \Big\{\hat{1} + s \hat{P}
\nonumber
\\
&+&
s^2 \,\hat{1} \Big[ R_{\mu \nu} f_1 (-s\Box) R^{\mu\nu}
+ R f_2(-s\Box) R
\nonumber
\\
&+&
\hat{P} f_3(-s\Box) R
+ \hat{P} f_4(-s\Box) \hat{P}
\nonumber
\\
&+&
\hat{S}_{\mu\nu} f_5 (-s\Box) \hat{S}^{\mu\nu} \Big] \Big\}\,,
\eeq
where  $\,\om\,$ is the parameter of dimensional regularization,
$\,\mu\,$ is renormalization parameter and the functions
$\,f_{1,2,...,5}\,$ are given by the following expressions:
\beq
f_1 (\tau) &=& \frac{f(\tau) - 1 + \tau/6}{\tau^2}\,,
\nonumber
\\
f_2 (\tau) &=& \frac{f(\tau)}{288} + \frac{f(\tau)-1}{24 \tau}
- \frac{f(\tau)-1+\tau/6}{8 \tau^2}\,,
\nonumber
\\
f_3 (\tau) &=& \frac{f(\tau)}{12} +
\frac{f(\tau)-1}{2\tau}\,,
\quad
f_4 (\tau) \,=\, \frac{f(\tau)}{2}\,,
\nonumber
\\
f_5 (\tau) &=&
\frac{1 - f(\tau)}{2\tau}\,,
\eeq
where
\beq
f(\tau) = \int_0^1 \, d\al \, e^{\al(1-\al)\tau}
\,, \qquad \tau = - s \Box \,.
\eeq

The integrals were taken in Ref.~\cite{apco, fervi} and we
present only the final result,
\beq
\n{0-EAM}
\bar{\Ga}^{(1)}
&=&
- \frac{1}{2(4 \pi)^2} \int d^{2\om} x \, \sqrt{g}\,
\Big\{ l_{CC}L_0 + l_RL_1 R
\nonumber
\\
&+&
\sum_{i=1}^5 l_i^* \,R_{\mu\nu} \,M_i \,R^{\mu\nu}
+ \sum_{i=1}^5 l_i \,R \,M_i \,R \Big\},
\eeq
where the coefficients $\,l^*_{1,2,..,5}\,$ and $\,l_{CC,R,1,2,..,5}\,$
are model-dependent and the integrals are universal. Within the
dimensional regularization $\,\om \to 2\,$ they have the form
\beq
\n{Int-L0}
L_0 &=& \frac{m^4}{2}\Big(\frac{1}{\epsilon}+\frac{3}{2}\Big)\,,
\\
L_1 &=&  -m^2 \Big(\frac{1}{\epsilon}  + 1 \Big)\,,
\label{Int-L1}
\\
M_1 &=&  \frac{1}{\epsilon} + 2 Y \,,
\label{Int-M1}
\\
M_2
&=&
\Big(\frac{1}{\epsilon} + 1 \Big)
\Big(\frac{1}{12}-\frac{1}{a^2}\Big)
- \frac{4Y}{3a^2}
+ \frac{1}{18} \,,
\label{Int-M2}
\\
M_3 &=&  \Big(\frac{1}{\epsilon} + \frac{3}{2}  \Big)
\Big(\frac{1}{2a^4}-\frac{1}{12 a^2}+\frac{1}{160}\Big)
\nonumber
\\
&+&
\frac{8Y}{15a^4}
-\frac{7}{180a^2}+\frac{1}{400}\,,
\label{Int-M3}
\\
M_4
&=&
\Big(\frac{1}{\ep} + 1 \Big)
\Big(\frac{1}{4} - \frac{1}{a^2} \Big)\,,
\label{Int-M4}
\\
M_5 &=&  \Big( \frac{1}{\epsilon} + \frac{3}{2} \Big)
\Big( \frac{1}{2a^4}-\frac{1}{4 a^2}+\frac{1}{32}\Big)\,,
\label{Int-M5}
\eeq
where we used condensed notation
\beq
\frac{1}{\epsilon} = \frac{1}{2-\om} - \ga
 + \ln \Big(\frac{4\pi \mu^2}{m^2}\Big)
\eeq
and $\ga$ is the Euler number (it was absorbed into
$\mu$-dependence in \cite{apco,fervi}).  In the expressions
\eq{Int-L0}-\eq{Int-M5} we disregarded the terms with are
${\cal O}(2-\om)$. We also used the definitions
\beq
\n{A}
Y
 &=&
1 - \frac{1}{a} \, \ln \Big(\frac{2 + a}{2 - a} \Big)
\eeq
and
\beq
\n{a-square}
a^2 = \frac{4 \Box}{ \Box - 4m^2}.
\eeq

In order to arrive at the final form of the one-loop effective
action, it is useful to introduce the basis which consists from
the square of the Weyl tensor and of scalar curvature. For this
end one can assume that for functions $\,F=F(\Box)$ of our interest
there is an expansion into power series in $\Box$, and use the
reduction formula (see, e.g., \cite{highderi})
\beq
R_{\mu\nu\al\be} F R^{\mu\nu\al\be}
&=&
4 R_{\mu\nu}  F R^{\mu\nu} - R F R
+
{\cal O}(R_{\dots}^3)\,.
\label{GB}
\eeq
For the scalar field with non-minimal interaction to external
gravity,
\beq
\n{S0}
S_0 = \frac{1}{2} \int d^4 x \sqrt{g}\,
\Big\{ g^{\mu\nu} \pa_\mu \ph \pa_\nu
\ph + m^2 \ph^2 + \xi R \ph^2 \Big\}\,,
\eeq
the ${\cal O}(R_{\dots}^2)$ result is\footnote{Here we
follow \cite{apco} and use an opposite sign for the mass term.
This is reasonable taking into account possible applications to
spontaneous symmetry breaking.}
\cite{apco}
\beq
\n{EA-scalar}
\bar{\Ga}_0^{(1)}
&=&
- \frac{1}{2(4 \pi)^2} \int d^4x \, \sqrt{g} \,
\Big\{ \frac{m^4}{2}\Big(\frac{1}{\ep}+\frac{3}{2}\Big)
\nonumber
\\
&+&
\tilde{\xi} \Big(\frac{1}{\epsilon}  + 1 \Big) m^2 R
+ R \Big(\frac{1}{2 \epsilon } \tilde{\xi}^2
+ k^0_R \Big) R
\nonumber
\\
&+&
\frac{1}{2} \, C_{\mu\nu\al\be} \Big(\frac{1}{60 \epsilon }
+ k^{0}_W \Big) C^{\mu\nu\al\be}
\Big\}\,,
\eeq
where $\,\tilde{\xi}=\xi - 1/6\,$ and
 the non-local form factors have the form
\beq
\n{Form-s1}
k^0_W \,=\,k^0_W (a) =
\frac{1}{150} + \frac{2}{45 a^2} + \frac{8 Y}{15 a^4}
\eeq
and
\beq
\n{Form-s2}
 k^0_R
 &=&
 k^0_R (a) \,=\,
 \frac{1}{108}\Big( \frac{1}{a^2}
- \frac{7}{20}\Big)
+ \frac{Y}{144}\Big(1- \frac{4}{a^2}\Big)^2
\nonumber
\\
&+&
\Big(\frac{1}{18}
- \frac{Y}{6}
+ \frac{2 Y}{3 a^2}\Big)\,\tilde{\xi}
+ Y \tilde{\xi}^2\,.
\eeq

In the next sections we will need the form factors for a minimal
(means $\xi=0$) scalar and also the ones for the Proca model
in curved space,
\beq
\n{ac-1}
S_1
&=&
\int d^4 x \sqrt{g} \,\Big\{ - \frac{1}{4}\, F_{\mu\nu}^2
- \frac12\, m^2 A_{\mu}^2 \Big\} \,,
\eeq
where $\,F_{\mu\nu} = \na_\mu A_\nu - \na_\nu A_\mu$.
The standard Stueckelberg procedure can be easily adapted to
the curved space \cite{BuGui},
yielding an equivalent action with an extra scalar  field $\,\ph$,
\beq
\n{New-ac1}
S'_1
 =
- \int d^4 x \sqrt{g} \Big\{
\frac{1}{4}\, F_{\mu\nu}^2
+ \frac{m^2 }{2} \big( A_{\mu}
- \frac{1}{m}\pa_\mu\ph\big)^2 \Big\} .
\eeq
The new action \eq{New-ac1} is gauge invariant under the gauge
transformations
\beq
A_\mu \to A'_\mu = A_\mu + \na_\mu f,
\quad
\ph \to \ph' = \ph + m f\,.
\label{trans}
\eeq
The original theory \eq{ac-1} is recovered in the special gauge
$\,\ph = 0$. Since the gauge fixing dependence is irrelevant for
the derivation of vacuum effective action, the practical calculation
can be performed in a more useful gauge. The reader can consult
Ref.~\cite{BuGui} for the details, let us just present the final result
\beq
\n{EA-vetor}
\bar{\Ga}^{(1)} &=&
- \frac{1}{2(4 \pi)^2} \int d^4x \, \sqrt{g} \,
\Big\{ \frac{3 m^4}{2} \Big(\frac{1}{\epsilon}
+\frac{3}{2}\Big)
\nonumber
\\
&+&
\frac{m^2}{2} \Big(\frac{1}{\epsilon}  + 1 \Big) R
+ R \Big(\frac{1}{72 \ep} + k^1_R  \Big) R
\nonumber
\\
&+&
\frac{1}{2} \, C_{\mu\nu\al\be} \Big(\frac{13}{60 \epsilon }
+ k^1_W \Big) C^{\mu\nu\al\be}\Big\}\,,
\eeq
where
\beq
k^1_W
&=&
k^1_W (a)
= -\frac{91}{450}
+\frac{2}{15 a^2}
+ Y
+\frac{8 Y}{5 a^4}
-\frac{8 Y}{3 a^2}\,,
\label{ff-Proca}
\\
k^1_R
&=&
k^1_R (a)
= -\frac{1}{2160}
+\frac{1}{36 a^2}
+\frac{Y}{48}
+\frac{Y}{3 a^4}-\frac{Y}{18 a^2}\,.
\nonumber
\eeq
As it was already discussed in \cite{BuGui}, the massless limit
in the expression (\ref{EA-vetor}) does not yield the effective
action for a massless field, due to the discontinuity in the
quantum corrections. In the next sections we shall meet two
other examples of a similar discontinuity in the massless limit.

\section{Massive antisymmetric rank-2 tensor}
\label{sec-rank2}

In this section we first present the well-known general considerations
and then proceed to the derivation of non-local form factors.

\subsection{General considerations}

The model of massive antisymmetric second-rank tensor
$\,B_{\mu\nu} = - B_{\nu\mu}\,$ field is described by the action
\beq
\n{Ant2-Act}
S_2
= \int d^4 x\sqrt{g}
\Big\{
- \frac{1}{12} F_{\mu\nu\la} F^{\mu\nu\la}
- \frac{m^2}4   B_{\mu\nu}B^{\mu\nu} \Big\},
\eeq
where
\beq
F_{\mu\nu\la}
&=& \na_\mu B_{\nu\la}
+ \na_\nu B_{\la \mu} + \na_\la B_{\mu\nu}\,.
\n{F}
\eeq
In four dimensions the theory \eq{Ant2-Act} is classically
equivalent to a massive axial vector field $A^\mu$. The
equivalence can be found through detailed analysis of the
equations of motion\footnote{
Let us note that the massive axial  vector describes an
antisymmetric torsion field. This identification comes from
the requirement of quantum consistency \cite{belyaev,torsi}.
The theory was eventually shown to violate unitarity when
coupled to fermions \cite{guhesh}.}. The duality between the
two theories is given by
\beq
B_{\mu\nu}
&  \propto  &
\frac{1}{m}\, \epsilon_{\mu\nu\al\be}
F^{\al\be}\,,
\eeq
where $\,F_{\mu\nu} = \na_\mu A_\nu - \na_\nu A_\mu$.

The model \eq{Ant2-Act} is an example of a theory with the
softly broken gauge symmetry. The kinetic part of the action
\eq{Ant2-Act} is gauge invariant under the transformation
\beq
\n{r2-gauge}
B_{\mu\nu} \to B'_{\mu\nu} = B_{\mu\nu}
+ \na_\mu \xi_\nu - \na_\nu \xi_\mu\,,
\eeq
where the vector gauge parameters $\xi_\mu$ in \eq{r2-gauge}
are not unique. They can be transformed according to
\beq
\n{xi-gauge}
\xi_\mu \to \xi_\mu' = \xi_\mu + \na_\mu \ph\,,
\eeq
with $\,\ph=\ph (x)\,$ being an arbitrary scalar field. Equation
\eq{xi-gauge} means that the gauge generators are linearly
dependent. Using the background field method we can observe
that the massive term in \eq{r2-gauge} violates gauge symmetry,
but does not remove the degeneracy in the bilinear form in
quantum fields
\beq
\hat{H}_2
&=&
\frac12\,
\frac{\de^2 S_2}{\de B_{\mu\nu}(x)\,\de B_{\al\be}(x^\prime)}\,.
\eeq

Similar to the Proca model,  the simplest approach for the
Lagrangian quantization of the theory (\ref{Ant2-Act}) requires
the Stueckelberg procedure. Following \cite{Buch-Kiri-Plet}, we
introduce an extra vector field $A_\mu$ and consider, instead
of Eq. \eq{r2-gauge}, the action
\beq
\n{Ant2-Act-New}
S'_2
&=&
\int d^4 x \, \sqrt{g} \,
\Big\{ - \frac{1}{12} \, F_{\mu\nu\la}
F^{\mu\nu\la}
\nonumber
\\
&-&
\frac{1}{4}\, m^2 \Big( B_{\mu\nu}
- \frac{1}{m} F_{\mu\nu}\Big)^2 \Big\}\,.
\eeq

The previous action \eq{Ant2-Act} can be obtained from
\eq{Ant2-Act-New} in the specific gauge $\,A_\mu= 0$.

The new action \eq{Ant2-Act-New} is gauge invariant under
the simultaneous transformation
\beq
\n{G1}
B_{\mu\nu} &\to& B'_{\mu\nu} = B_{\mu\nu}
+ \na_\mu \xi_\nu - \na_\nu \xi_\mu\,,
\\
\n{G2}
A_\mu &\to& A'_\mu = A_\mu + m \, \xi_\mu \,
\eeq
and it is also invariant under gauge transformation of the
Stueckelberg field
\beq
\n{G3}
A_\mu \to A'_\mu = A_\mu +\na_\mu \La \,,
\eeq
with a scalar parameter $\La(x)$.
Furthermore, we can consider a new scalar field $\ph(x)$
and note that the fields $B_{\mu\nu}$ and $A_\mu$ do not
change if their gauge parameters transform as
\beq
\n{D1}
\xi &\to& \xi_\mu' = \xi_\mu + \na_\mu \ph \,,
\\
\n{D2}
\La &\to& \La' = \La + m \ph \,.
\eeq
Once again, the equations \eq{D1} and \eq{D2} reflect the  fact
that the gauge generators of the theory are linearly dependent.

The general formalism of Lagrangian quantization in theories with
dependent generators is based on the Batalin-Vilkovisky method
\cite{Batalin-Vilkovisky}. However, in the relatively simple theories
such as \eq{Ant2-Act-New}, where the action is quadratic and the
algebra of dependent gauge generators is Abelian, it is sufficient
to make a successive multi-step application of the Faddeev-Popov
method \cite{Schwarz1,Schwarz2,13}. In the
following we are going apply this approach to the  theory
\eq{Ant2-Act-New}.

According to the Faddeev-Popov method we replace the gauge
group integration in the functional integral over gauge fields,
\beq
\int DB \, DA \, e^{\,i\, S'_2[B,A]}\,,
\label{base}
\eeq
by the quantity
\beq
\n{Standard-FP}
\int DB DA  e^{i\, S'_2[B,A]} \, \De\,
\de \big(\chi^\al[B,A] - l^\al ,\chi[A] - l \big),
\eeq
where $\De$ provides the identity
\beq
\n{Standard-De}
1 = \De  \int D \xi D \La \,
\de \left(\chi^\al[B',A'] - l^\al ,\,\chi[A'] - l \right).
\eeq
Here $\chi^\al$ and $\chi$ are the gauge fixing term
which are related to the transformations \eq{G1}, \eq{G2}
and to \eq{G3}, respectively. For the theory \eq{Ant2-Act-New}
one can choose
\beq
\chi_{\be}
&=&
\na^\al B_{\al\be} - m \, A_\be \,,
\qquad
\chi \,=\, \na_\al A^\al \,.
\eeq
It is easy to verify the following constraint between the two gauge
fixing conditions:
\beq
\n{Transverse}
\na^\al \chi_\al + m \chi = 0 \,.
\eeq

Due to the constraint \eq{Transverse}, the delta-function
$\,\de \left(\chi^\al[B,A] ,\, \chi[B] \right)\,$ in the definitions
\eq{Standard-FP} and \eq{Standard-De} is ill-defined, that
represents the main difference with the standard Faddeev-Popov
procedure.  The same problem also affects the integral in
\eq{Standard-De}, since the fields are invariant under
transformations \eq{D1} and \eq{D2}
for the gauge parameters $\xi_\mu$ and $\La$ . To solve this
issue, one can also apply the Faddeev-Popov trick second time
to remove the integration along the gauge group orbits.
Consider the Fourier representation for the delta-function
\beq
\n{DeDirac-int-1}
\de \left(\chi^\al ,\, \chi \right)
&=&
\int \, D \zeta \, D \psi
\, e^{\,i(\, \ze_\al \chi^\al\, -\,\psi\,\chi)} \,.
\eeq
After integration by parts and using \eq{Transverse} one can
easily show that the integrand in \eq{DeDirac-int-1} is invariant
under the transformation
\beq
\n{DD1}
\ze_\al &\to& \ze_\al' = \ze_\al + \na_\al \phi \,,
\\
\n{DD2}
\psi &\to& \psi' = \psi + m\, \phi \,.
\eeq
In order to have well-defined definition one can extract
from  \eq{DeDirac-int-1} the integral over gauge group orbit
\eq{DD1}-\eq{DD2} by using the Faddeev-Popov trick,
hence we arrive at
\beq
\n{DeDirac-int-2}
\de \left(\chi^\al \,,\, \chi \right)
&=&
\int \, D \zeta \, D \psi  \, e^{\,
i( \zeta_\al \, \chi^\al - \, \psi \, \chi) } \, \times
\nonumber
\\
&\times & \de ( \na_\al \ze^\al - m\psi)
\,\Det \hat{H}_0^{min} \,,
\eeq
where $\,H_0^{min}= \Box - m^2\,$ is a minimal scalar operator.
Let us remember that this operator depends on the external metric
and its functional determinant is non-trivial.

For the integral in Eq.~\eq{Standard-De} one has to factorize
the integrations over gauge group orbits \eq{D1}-\eq{D2}.
This means we replace $\,D \xi D\La\,$ by the product
\beq
D \xi D\La \, \de (\na_\al \xi^\al - m\La) \, \Det \hat{H}_0^{min}
\eeq
 in the definition \eq{Standard-De}. Hence, the equation
 \eq{Standard-De} for $\De$ becomes
\beq
\De^{-1} &=& \int \, D \ze \, D\psi \, D \xi \, D \La  \, \de (\na_\al
\ze^\al - m\psi - s )\, \times
\nonumber
\\
&\times &  \de (\na_\be \xi^\be - m\La - t )\,\times
\n{De-certo}
\\
&\times& e^{i \, \left\{\ze_\al \, \left(\chi^\al[B'] - l^\al \right)
- \psi \, (\chi[A'] - l) \right\}} \,
( \Det  \hat{H}_0^{min})^2   \,.
\nonumber
\eeq
For solving \eq{De-certo} one can use the fact that
\beq
\n{vs}
\chi ^\al [B',A'] - l^\al
&=& \chi ^\al [B,A] - l^\al + (H_1)^\al_\be
\,\xi^\be
\nonumber
\\
\mbox{and}
\quad
\chi[A'] - l &=& \chi[A] - l + \Box \La \,,
\eeq
where the vector operator is
\beq
\n{Bilinear-1}
\hat{H}_1 &=&
(H_1)_{\nu}^{\mu}
\,=\,
\de^\mu_\nu \Box - \na^\mu \na_\nu - R^\mu_\nu
- m^2 \de^\mu_\nu \,.
\eeq

Because of the delta-function in \eq{Standard-FP}, the fields
satisfy the equations
$\,\chi[B,A]^\al-l^\al =0\,$ and $\,\chi[A]^\al-l =0$.
Therefore, introducing the identity factor in the form of the double
integral $\,\int Ds \, Dt \, e^{- i s\, t}\equiv 1$, we can take the
integral over delta-functions, arriving at
\beq
\n{De-Re}
\De = \Det \hat{H}'_1 \,\cdot\,
(\Det \hat{H}_0^{min})^{-1}
\,,
\eeq
with $\,\hat{H}'_1\,$ is a minimal vector operator,
\beq
\n{H1prime}
\hat{H}'_1 = (H_1^\prime)_\nu^\mu =
\de^\mu_\nu \Box - R^\mu_\nu - \de^\mu_\nu m^2 \,.
\eeq
For the sake of completeness, let us remember the Stueckelberg
procedure for the massive vector field \cite{BuGui},
\beq
\n{EA-1}
\bar{\Ga}^{(1)} &=&  \frac{1}{2} \Tr \ln \hat{H}_1
\nonumber
\\
&=&
 \frac{1}{2}
\Tr \ln ( \de^\mu_\nu \Box - R^\mu_\nu - \de^\mu_\nu m^2)
- \frac{1}{2} \Tr \ln (\Box - m^2)
\nonumber
\\
&=& \frac{1}{2} \Tr \ln \hat{H}'_1
- \frac{1}{2} \Tr \ln \hat{H}_0^{min} \,.
\eeq

Now, using the well-defined expressions \eq{DeDirac-int-2} and
\eq{De-Re}  we can consider  Eq.~\eq{Standard-FP} and then
the effective  action. First one has to write the delta-function
\eq{DeDirac-int-2} in a  more useful way. Using the Fourier
representation
\beq
\de (\na_\al \ze^\al - m \psi)
&=&
\int D \ph \, e^{i\,(\na_\al \ze^\al - m\psi) \, \ph}
\nonumber
\\
&=&
\int D \ph \, e^{i\, ( - \ze^\al \na_\al \ph  - \psi \, m \, \ph )} \,,
\eeq
we can make an integration over $\ze$ and $\psi$ in
\eq{DeDirac-int-2} and find
\beq
\n{De-final}
\de \left(\chi^\al  ,\, \chi  \right)
&=& \int  D \ph \,\, \de
\left(\chi_\al - \na_\al \ph \right) \,\times
\nonumber
\\
&\times &
\de \left(\chi + m \ph \right) \, \Det \hat{H}_0^{min}.
\eeq
Hence, using Eqs.~\eq{De-final} and \eq{De-Re} we write
the vacuum effective action in the form
\beq
\n{VP-EA-1}
e^{i \Ga[g_{\mu\nu}]} &=& \int DB \, DA \, D \ph
\, e^{\,i\, S'_2[B,A]} \,
\de \left(\chi_\al - \na_\al \ph -l^\al \right) \,\times
\nonumber
\\
&\times&
\de \left(\chi + m \ph - l \right) \,
\Det \hat{H}'_1 \,.
\eeq
As far as \eq{VP-EA-1} does not depend of $l^\al$ and $l$ one can
insert the identities in the form of
$\,\int Dl \, e^{-\frac{i}{2} l^\al l_\al}$
and $\int Dl \, e^{-\frac{i}{2} l^2}$ to take the integrals. In this
way we arrive at
\beq
e^{i \Ga[g_{\mu\nu}]}
&=&
\int DB \, DA \, D \ph \,
e^{i\left\{S'_2[B,A] + S_{gf}[B,A] - S_0[\ph] \right\}} \,\times
\nonumber
\\
&\times&
\Det \hat{H}'_1 \,,
\eeq
where $S_{gf} = -\frac{1}{2} \int d^4 x \sqrt{g} \{\chi_\al \chi^\al +
\chi^2\}$ and $S_0[\ph]$ is the action of scalar field (\ref{S0}).

The action with the gauge-fixing term can be written in the form
\beq
\n{Bili2}
S'_2 + S_{gf}
&=&
\int d^4 x \sqrt{g} \Big\{ \frac{1}{4} \, B_{\al\be}
(H'_2)^{\mu\nu}_{\al\be} B^{\mu\nu}
\nonumber
\\
&+&
\frac{1}{2} \, A_\mu (H'_1)^\mu_\nu A^\nu \Big\} \,,
\eeq
where
\beq
\hat{H}'_2
&=&
(H'_2)^{\al\be}_{\quad\mu\nu}
\nonumber
\\
&=&
 \de^{\al\be}_{\quad\mu\nu} (\Box - m^2)
- J^{\al\be}_{\quad\mu \nu}
+ R^{\al\be}_{\,.\,.\,\mu\nu}
\eeq
and $\,(H'_1)^\mu_\nu\,$ was defined in (\ref{H1prime}).
Here
\beq
\de_{\al \be ,\, \mu \nu} = \frac{1}{2} \left(
g_{\al\mu} g_{\be\nu} - g_{\al\nu}g_{\be\mu}
\right) ,
\eeq
is the identity matrix in the antisymmetric rank-2 tensor
space and
\beq
J_{\al \be ,\, \mu \nu} = \frac{1}{2} \, \big(
\,g_{\al\mu} R_{\be\nu}
+ g_{\be\nu} R_{\al\mu}
- g_{\al\nu} R_{\be\mu}
- g_{\be\mu} R_{\al\nu}\, \big)\,.
\nonumber
\eeq

Eqs.~\eq{VP-EA-1} and \eq{Bili2} enable one to formulate the one-loop
contribution to the vacuum Euclidean effective action in the form
\beq
\n{EAe}
\bar{\Ga}^{(1)}
&=&   \frac{1}{2}( \Tr \ln \hat{H}'_2 + \Tr \ln \hat{H}'_1
+ \Tr \ln \hat{H}_0^{min} ) - \Tr \ln \hat{H}'_1
\nonumber
\\
&=&  \frac{1}{2} \Tr \ln \hat{H}'_2  - \frac{1}{2} \Tr \ln \hat{H}'_1
+ \frac{1}{2} \Tr \ln \hat{H}_0^{min}\,.
\eeq

Using the relation for the Proca field contribution, Eq. ~(\ref{EA-1}),
one  can also write the one-loop effective action in the form
\beq
\bar{{\Ga}}^{(1)}
&=&
\frac{1}{2} \Tr \ln \hat{H}'_2
- \frac{1}{2} \Tr \ln \hat{H}_1 \,.
\label{Gamma}
\eeq
Finally, the effective action requires subtracting the contribution
of the Stueckelberg  massive vector from the one of the massive
tensor operator, $\,\Tr \ln\hat{H}'_2$.

\subsection{Derivation of form factors}

The result (\ref{EAe})  enables one to use the heat-kernel
technique for deriving form factors.  The first step is to
identify the elements of the general expression (\ref{0-EAM}),
\beq
\hat {1}
&=&
\de^{\al\be}_{\quad\mu\nu}\,,
\nonumber
\\
\hat{P}_2 &=&
(P_2)^{\al\be}_{\quad\mu\nu} = R^{\al \be}_{\,.\,.\,\mu \nu}
+ \frac{1}{6} \, \de^{\al \be}_{\quad\mu \nu} R
- J^{\al \be}_{\quad\mu \nu}\,.
\eeq
The commutator of covariant derivatives on the antisymmetric
tensor field $B_{\mu\nu}$ is
\beq
&&
\hat{(S_2)}_{\mu\nu}
\,=\,
\big[(S_2)_{\mu\nu}\big]^{\al\be}_{\rho\om}
\\
\,=\,
&& \frac{1}{2} \,\big(R^{\al}_{.\,\rho\mu\nu} \, \de^\be_\om
- R^{\be}_{.\,\rho\mu\nu} \, \de^\al_\om
- R^{\al}_{.\,\om\mu\nu} \, \de^\be_\rho
+ R^{\be}_{.\,\om\mu\nu} \, \de^\al_\rho\big)\,.
\nonumber
\eeq
Then, using the heat kernel representation, we arrive at the
identification in the second order in curvatures,
\beq
&& \frac{1}{2} \Tr \ln H'_2
\,=\,
- \frac{1}{2(4 \pi)^2} \int d^{2\om} x \, \sqrt{g} \,
\Big\{\, 6 L_0 - L_1 \, R
\nonumber
\\
&&
+ \sum_{i=0}^5 l_i^* \,R_{\mu\nu} \,M_i \,R^{\mu\nu}
+ \sum_{i=0}^5 l_i \,R \,M_i \,R \, \Big\}\,,
\label{hk}
\eeq
where $l_{CC}$ and $l_R$ are already inserted into (\ref{hk}),
other coefficients are
\beq
l_1 = -\frac{5}{16}
\,,
\quad
l_2 = -\frac{5}{4}\,,
\quad
l_3 = -\frac{3}{4}\,,
\quad
l_4 = \frac{9}{8}\,,
\quad
l_5 = \frac{3}{4}\,,
\nonumber
\eeq
\beq
l_1^* = 1\,,
\quad
l_2^* = 4\,,
\quad
l_3^* = 6\,,
\quad
l_4^* = -3\,,
\quad
l_5^* = -6\,.
\nonumber
\eeq
Replacing these values and the integrals \eq{Int-L0}-\eq{Int-M5},
we arrive at the expression
\beq
\n{EA-2'}
\frac{1}{2} \, \Tr \ln \hat{H}'_2 &=&
- \frac{1}{2(4 \pi)^2} \int d^4x \, \sqrt{g} \,
\Big\{
3 m^4 \Big(\frac{1}{\ep}+\frac{3}{2}\Big)
\nonumber
\\
&+&
m^2 \Big(\frac{1}{\ep}  + 1 \Big)  R
+ R \Big[\frac{1}{36 \ep}
+ k'_{2,R} (a) \Big] R
\nonumber
\\
&+& \frac{1}{2} \, C_{\mu\nu\al\be} \Big[\frac{13}{30 \ep}
+ k'_{2,W} (a) \Big] C^{\mu\nu\al\be}
 \Big\}\,,
\eeq
where
\beq
k'_{2,W} (a) =
-\frac{91}{225}
+\frac{4}{15 a^2}
+2 Y
+\frac{16 Y}{5 a^4}
-\frac{16 Y}{3 a^2}
\,,
\eeq
\beq
k'_{2,R} (a) =
-\frac{1}{1080}
+\frac{1}{18 a^2}
+\frac{Y}{24}
+\frac{2 Y}{3 a^4}
-\frac{Y}{9 a^2}
\,.
\eeq
According to Eq.~(\ref{Gamma}), we have to subtract from
\eq{EA-2'} the massive vector part, Eq.~\eq{EA-vetor}.
Hence, we get
\beq
\n{EA-2a}
\bar{\Ga}^{(1)}
&=&
-\,\frac{1}{2(4 \pi)^2} \int d^4x \, \sqrt{g} \,
\Big\{ \frac{3 m^4}{2}\,\Big(\frac{1}{\ep}
+\frac{3}{2}\Big)
\nonumber
\\
&+&
\frac{m^2}{2}\, \Big(\frac{1}{\ep}  + 1 \Big) R
+ R \Big[\frac{1}{72 \ep} + k^2_R \Big] R
\nonumber
\\
&+&
\frac{1}{2} \, C_{\mu\nu\al\be} \Big[\frac{13}{60 \ep}
+ k^2_W \Big] C^{\mu\nu\al\be}\Big\}\,,
\eeq
where the non-local form factors are
\beq
k^2_W  &=& k^2_W(a) =
-\frac{91}{450}
+\frac{2}{15 a^2} +Y +\frac{8 Y}{5 a^4} -\frac{8 Y}{3 a^2}\,,
\\
k^2_R &=& k^2_R(a) =
-\frac{1}{2160}
+\frac{1}{36 a^2}
+\frac{Y}{48}
+\frac{Y}{3 a^4}-\frac{Y}{18 a^2}\,.
\nonumber
\eeq
It is easy to see that the vacuum effective action for the massive
rank-2 antisymmetric tensor, \eq{EA-2a}, is exactly the same as
the vacuum effective action in the massive vector field case,
 given by Eq.~\eq{EA-vetor}. This confirms the conclusion of
 \cite{Buch-Kiri-Plet} that the massive rank-$2$ antisymmetric
 tensor is equivalent to the Proca theory at quantum level. Let us
 note that this conclusion has been achieved by the
 $\ze$-regularization method, and we know that some of
 the relations of this kind can be violated by the non-local
 multiplicative  anomaly \cite{QED-form}. Nothing  of this sort
 occurs in the present case, as we have seen.

It is easy to check that in the massless limit the form factor
$k^2_W(a)$ reduce to the usual logarithmic expression
$\,-\frac{13}{60} \ln \left( - \frac{  \Box}{4 \pi \mu^2}\right)$.
On the other hand, we know that in the $m=0$ case the rank-$2$
antisymmetric tensor is equivalent to a scalar field minimally
coupled to gravity,  where the duality looks like
$\,F_{\al\be\om} = \ep_{\al\be\om\ga} \na^\ga \ph$.
The form factor for the minimal massless scalar field is
 $\,-\frac{1}{60} \ln \left( - \frac{\Box}{4 \pi \mu^2} \right)$.
 The difference between the two coefficients
$1/5 = 13/60 - 1/60$ demonstrates the discontinuity of quantum
contributions in the massless limit for the rank-$2$ antisymmetric
tensor theory.
This difference is nothing else but the contribution of a massless
vector field. To understand which vector is this, let us consider
the effective action for a massless rank-$2$ antisymmetric
tensor field \cite{Buch-Kiri-Plet},
\beq
\n{EA2m}
\bar{\Ga}^{(1)} =  \frac{1}{2} \Tr \ln \hat{H}'_2  - \Tr \ln \hat{H}'_1
+ \frac{3}{2} \Tr \ln \hat{H}_0^{min}\,.
\eeq
The difference between \eq{EAe} and \eq{EA2m} is
\beq
-1/2 \Tr \ln \hat{H}'_1 +\Tr \ln \hat{H}_0^{min} ,
\eeq
which is the effective action for the free massless vector field.
Another way to understand this is to recall that massive rank-$2$
antisymmetric tensor field is equivalent to a massive vector field
model. According to \cite{BuGui},
\beq
\n{EAe2}
\bar{\Ga}^{(1)}
&=&
 \frac{1}{2} \Tr \ln ( \de^\mu_\nu \Box - R^\mu_\nu
- \de^\mu_\nu m^2)
+  \frac{1}{2} \Tr \ln (\Box - m^2)
\nonumber
\\
&-&
\Tr \ln (\Box - m^2)
\\
&=&
\frac{1}{2} \Tr \ln ( \de^\mu_\nu \Box - R^\mu_\nu
- \de^\mu_\nu m^2)
\,-\, \frac{1}{2} \Tr \ln (\Box - m^2) \,.
\nonumber
\eeq
Obviously, the difference in the contributions of a massive vector
field  and the one of the minimal scalar field is just the effective
action of a  massless vector field. In the massless limit this extra
term does  not disappear and this produce the quantum
discontinuity.

\section{Massive antisymmetric rank-3 tensor}
\label{sec-rank3}

As a second example, consider the model of massive totally
antisymmetric rank-$3$ tensor field
$\,C_{\mu \nu \rho} = C_{[\mu \nu \rho]} $.
The action is given by
\beq
\n{S3}
S_3
&=&
\int d^4 x\,\sqrt{g}\,\Big\{
- \frac{1}{48} \, F_{\mu\nu\rho\om} ^2
- \frac{1}{12} \, m^2\,  C_{\mu\nu\rho} ^2
\Big\},
\eeq
where
\beq
F_{\mu\nu\rho\om}
=
\na_\mu C_{\nu\rho\om} - \na_\nu C_{\rho\om\mu}
+ \na_\rho C_{\om\mu\nu} - \na_\om C_{\mu\nu\rho} .
\nonumber
\eeq
It is possible to prove that in four dimensional space the theory
\eq{S3} is classically equivalent to the theory of a real massive
scalar field $\,\ph\,$ minimally coupled to gravity. The duality
relation between the two theories is defined by the relation
$\,C_{\mu\nu\rho}
\propto \frac{1}{m}\, \epsilon_{\mu\nu\al\be} \na^\be \ph$.

The kinetic term of the action \eq{S3} is invariant under the gauge
transformations
\beq
\n{GGG1}
C_{\mu\nu\rho}
&\to&
C'_{\mu\nu\rho}
\nonumber
\\
&= &
C_{\mu\nu\rho}
+ \na_\mu \om_{\nu \rho} + \na_\nu \om_{\rho \mu}
+ \na_\rho \om_{\mu \nu}\,,
\eeq
with an antisymmetric tensor field parameter
$\,\om_{\mu\nu} = - \om_{\nu\mu}$.
This parameter is defined up to the gauge transformation
\beq
\n{LD1}
\om_{\mu \nu} \to \om'_{\mu \nu}
&=& \om_{\mu \nu} + \na_\mu \ze_\nu - \na_\nu \ze_\mu\,,
\eeq
where $\ze_\mu$ is a vector gauge field parameter. Furthermore,
$\ze_\mu$ is also defined up to the gauge transformation
\beq
\n{LD2}
\ze_\mu
&\to&
\ze'_\mu = \ze_\mu + \na_\mu \phi\,,
\eeq
with the scalar  field parameter $\,\phi (x)$.
Equations \eq{LD1} and \eq{LD2} mean that the gauge generators are
linearly dependent. As in the previous case of the second-rank tensor,
due to the gauge invariance of $F_{\mu\nu\rho\om}^2$ we have to
deal with a theory with softly broken gauge symmetry. Therefore, the
quantization must be done with the Stueckelberg procedure.

One can restore the gauge symmetric by introducing an extra
second-rank antisymmetric field $B_{\mu\nu}$. Consider the
following action:
\beq
\n{S3'}
S'_3
&=&
\int d^4 x \,   \sqrt{g} \,   \Big\{
- \frac{1}{48} \, F_{\mu\nu\rho\om} ^2
\nonumber
\\
& - &
\frac{1}{12} \, m^2\,
\Big( C_{\mu\nu\rho} - \frac{1}{m} F_{\mu\nu\rho}\Big)^2\Big\}\,,
\eeq
where $F_{\mu\nu\rho}$ is defined in (\ref{F}).
The action \eq{S3'} is gauge invariant under the simultaneous
transformation \eq{GGG1} and
\beq
B_{\mu\nu} \to B'_{\mu\nu} = B_{\mu\nu} + m \, \om_{\mu\nu}
\,.
\eeq
It is also invariant under the gauge transformation of  Stueckelberg
procedure field
\beq
B_{\mu\nu} \to B'_{\mu\nu} = B_{\mu\nu} + \na_\mu \xi_\nu - \na_\nu \xi_\mu
\,,
\eeq
where $\xi_\mu$ is a vector gauge parameter, defined up to a gauge
transformation $\xi'_\mu = \xi_\mu + \na_\mu \ph$ with a scalar
parameter $\,\ph(x)$. Since the gauge generators of the theory are
linearly dependent,  the quantization of the theory \eq{S3'} differs
from standard scheme and can be done in a way similar to the one
described above for the second-rank field. The successive multi-step
applications of Faddeev-Popov method for the antisymmetric rank-3
tensor is somehow more tedious than for the antisymmetric rank-2
field, hence we are not going to  bother the reader with the details
and present only the final formula for the one-loop effective action,
\beq
\n{EA-3}
\bar{\Ga}^{(1)} &=&
\frac{1}{2} \Tr \ln \hat{H}'_3
- \frac{1}{2} \Tr \ln \hat{H}'_2
\nonumber
\\
&+ &
\frac{1}{2} \Tr \ln \hat{H}'_1
- \frac{1}{2} \Tr \ln \hat{H}_0^{min}
\nonumber
\\
&=&  \frac{1}{2} \Tr \ln \hat{H}'_3
- \frac{1}{2} \Tr \ln \hat{H}_2 \,,
\eeq
where
\beq
\hat{H}'_3
&=&
(H'_3)^{\al\be\om}_{\mu\nu\rho}
\nonumber
\\
&= &
 \de^{\al\be\om}_{\mu\nu\rho} (\Box - m^2)
+ K^{\al\be\om}_{\mu\nu\rho} - L^{\al\be\om}_{\mu\nu\rho}\,,
\eeq
with
\beq
\n{de3}
\de^{\al\be\om}_{\mu\nu\rho} =
\frac{1}{6} \, \ep^{\al\be\om\si} \ep_{\mu\nu\rho\si}
=
\begin{vmatrix}
\de^\al_\mu & \de^\al_\nu & \de^\al_\rho \\
\de^\be_\mu & \de^\be_\nu & \de^\be_\rho \\
\de^\om_\mu & \de^\om_\nu & \de^\om_\rho \\
\end{vmatrix},
\eeq
\beq
\n{K2}
K^{\al\be\om}_{\mu\nu\rho} = 3 \, \de^{\al\be\om}_{\ga \theta \ph} \,
\de^{\ga\la\tau}_{\mu\nu\rho} \, R^{\theta\ph}_{.\,.\,\la\tau}
\eeq
and
\beq
\n{L}
L^{\al\be\om}_{\mu\nu\rho} =
\de^{\al\be\om}_{\ga \theta \ph} \left(
R^\ga_\la \, \de^{\la\theta\ph}_{\mu\nu\rho}
+  R^\theta_\la \, \de^{\la\ph\ga}_{\mu\nu\rho}
+  R^\ph_\la \,  \de^{\la\ga\theta}_{\mu\nu\rho} \right) .
\nonumber
\\
\eeq
Equation \eq{de3} defines the generalized Kronecker delta which
serves as an identity matrix in the space of third-order totally
antisymmetric tensors. It also has the property
\beq
\n{Iden}
\de^{\al\be\om}_{\mu\nu\rho} \, T_{\al\be\om}
&=&
T_{[\mu\nu\rho]}\,.
\eeq
Due to the identity \eq{Iden} one can write the expressions
\eq{K2} and \eq{L} in a compact way respecting their symmetries.
From the technical side, by using the definition  \eq{de3},
calculation of divergences can be mainly reduced to contractions
of the products of Levi-Civita symbols.

In accordance to the formula \eq{EA-3} we need to work with
the third-rank tensor and subtract the second-rank contribution
which is already known. Consider the field  strengths for the
first term,
\beq
\hat{P}_3 = (P_3)^{\al\be\om}_{\mu\nu\rho}
= K^{\al\be\om}_{\mu\nu\rho}
+ \frac{1}{6} \, \de^{\al\be\om}_{\mu\nu\rho} R
- L^{\al\be\om}_{\mu\nu\rho}
\,,
\eeq
\beq
(\hat{S}_3)_{\mu\nu} &=& [(S_3)_{\mu\nu}]^{\al\be\om}_{\theta\ph\ga}
\\
&=&
 \de^{\al\be\om}_{\eta\xi\ze}(
R^{\eta}_{.\,\la\mu\nu} \de^{\la \xi\ze}_{\theta\ph\ga}
+ R^{\xi}_{.\,\la\mu\nu} \de^{\la \ze\eta}_{\theta\ph\ga}
+ R^{\ze}_{.\,\la\mu\nu} \de^{\la \eta\xi}_{\theta\ph\ga}) \,.
\nonumber
\eeq
Then it is easy to obtain the expression
\beq
&&
\frac{1}{2} \, \Tr \ln \hat{H}'_3
= - \frac{1}{2(4 \pi)^2} \int d^{2\om} x \, \sqrt{g} \,
\Big\{\,
4 L_0 - \frac{1}{3} L_1 \, R
\nonumber
\\
&&
+\, \sum_{i=0}^5 l_i^* \,R_{\mu\nu} \,M_i \,R^{\mu\nu}
\,+\,\sum_{i=0}^5 l_i \,R \,M_i \,R \, \Big\}\,,
\eeq
where
\beq
&&
l_1 = - \frac{1}{8},
\quad
l_2 = l_3 = -\frac{1}{2},
\quad
l_4 = \frac{5}{12},
\quad
l_5 = \frac{1}{2};
\nonumber
\\
&&
l_1^* = \frac{1}{2},
\quad
l_2^* = 2,
\quad
l_3^* = 4,
\quad
l_4^* = -\frac{4}{3},
\quad
l_5^* = -4.
\nonumber
\eeq
By using the table of integrals \eq{Int-L0}-\eq{Int-M5} we find
\beq
\n{EA-3'}
\frac{1}{2} \, \Tr \ln \hat{H}'_3
&=& -\frac{1}{2(4 \pi)^2} \int d^4x \, \sqrt{g} \,
\Big\{
2 m^4 \Big(\frac{1}{\epsilon}+\frac{3}{2}\Big)
\nonumber
\\
&+&
\frac{m^2}{3}  \Big(\frac{1}{\epsilon}  + 1 \Big)  R
+ R \Big[\frac{1}{36 \ep} + k'_{3,R} (a) \Big] R
\nonumber
\\
&+& \frac{1}{2} \, C_{\mu\nu\al\be} \Big[\frac{7}{30 \epsilon }
+ k'_{3,W} (a) \Big] C^{\mu\nu\al\be} \Big\}\,,
\eeq
where
\beq
k'_{3,W} (a) =
-\frac{44}{225}
+\frac{8}{45 a^2}
+Y
+\frac{32 Y}{15 a^4}
-\frac{8 Y}{3 a^2}
\,,
\eeq
\beq
k'_{3,R} (a) =
-\frac{7}{540}
+\frac{1}{27 a^2}
+\frac{Y}{12}
+\frac{4 Y}{9 a^4}
-\frac{2 Y}{9 a^2}\,.
\eeq
Subtracting \eq{EA-2a} from \eq{EA-3'} we arrive at the final result
\beq
\n{EA-3a}
\bar{\Ga}^{(1)} &=&
- \frac{1}{2(4 \pi)^2} \int d^4x \, \sqrt{g} \,
\Big\{ \frac{ m^4}{2}\Big(\frac{1}{\epsilon}+\frac{3}{2}\Big)
\nonumber
\\
&-&
\frac{m^2}{6} \Big(\frac{1}{\epsilon}  + 1 \Big) R
+ R \Big(\frac{1}{72 \epsilon }
+ k^3_R \Big) R
\nonumber
\\
&+&
\frac{1}{2} \, C_{\mu\nu\al\be} \Big(\frac{1}{60\ep}
+ k^3_W  \Big) C^{\mu\nu\al\be}
\Big\}\,,
\eeq
where
\beq
k^3_W
&=&
k^3_W (a) = \frac{1}{150}
+\frac{2}{45 a^2}
+\frac{8 Y}{15 a^4}\,,
\\
k^3_R
&=&
k^3_R(a) =
-\frac{1}{80}
+\frac{1}{108 a^2}
+\frac{Y}{16}
+\frac{Y}{9 a^4}
-\frac{Y}{6 a^2}\,.
\nonumber
\eeq
In accordance to the general proof of \cite{Buch-Kiri-Plet},
the effective action for the massive rank-3 antisymmetric tensor
\eq{EA-3a} is exactly the same as the one for the massive scalar field
minimally coupled  to gravity, given by \eq{EA-scalar} with $\xi=0$.
There is no anomaly in the non-local part of effective action in this
case.

Consider the massless limit for the form factor $k_W^3(a)$
of the Weyl-squared term. Taking the $\,m \to 0\,$ limit  in
Eq.~\eq{EA-3a} we
meet a non-zero contribution in the form
$\,-\frac{1}{60} \ln \left(- \frac{\Box}{4 \pi \mu^2} \right)$.
At the same time, for $\,m = 0\,$ the rank-3 antisymmetric
tensor has no degrees of freedom and the result of the calculation
is different. By using the methods explained in section
\ref{sec-rank2}, we arrive at the expression
\beq
\n{EA3m}
\bar{\Ga}^{(1)}
&=&
\frac{1}{2} \, \Tr \ln \hat{H}'_3
- \Tr \ln \hat{H}'_2
\nonumber
\\
&+&
\frac{3}{2}\Tr \ln \hat{H}'_1
- 2 \Tr \ln \hat{H}_0^{min}\,.
\eeq
Using previous results it is easy to check that the equation \eq{EA3m}
in the massless case gives $\,\bar{\Ga}^{(1)} = 0$.  The difference
in the coefficients of the logarithmic form factors in $k_W^3(a)$
of the massless limit in a massive theory and of the strictly massless
case is $1/60$. A similar situation holds for the factor $k_R^3(a)$.
In the $\,m \to 0\,$ limit of a massive theory of the third-rank tensor
the logarithmic coefficient in the form factor  follows the divergent
term and we find the quantum contribution
$\,-\frac{1}{36}\ln \left(- \frac{\Box}{4 \pi \mu^2} \right)\,$
for the $R^2$-term. In the strictly massless case as the effective
action \eq{EA3m} vanishes and
we meet no contribution. The difference between the two
coefficients is $1/36$ and represents the minimal scalar field
contribution which do not disappear in the $m \to 0$ limit.
This example once again demonstrates the quantum discontinuity
for the massless limit.

Finally let us note that the conformal anomaly for vector and
scalar massless fields can not be reproduced within the dual
antisymmetric theories. The reason is that the duality takes
place only in the massive theories and in the massless limits
there is a discontinuity which makes reproduction of anomaly
impossible. In the strictly massless theories we checked that
the second-rank model does not possess the local conformal
symmetry. This is a natural result due to the equivalence
with a minimal (and hence non-conformal) scalar model.

\section{Conclusions}
\label{Conc}

By making direct calculations we have confirmed that the result of
the paper \cite{Buch-Kiri-Plet} concerning the quantum equivalence
of massive tensor fields of the second- and third-rank with vector
and scalar models holds in the non-local form factors.
Furthermore, one meets the discontinuity of the massless limits
of quantum contributions in both cases. In fact, for the massless
cases the mentioned equivalence does not hold. In particular, for
the rank-3 tensor field in the massless case there is neither classical
nor  quantum dynamics and the theory is trivial.
It would be interesting to formulate the same two types of
fields on a more general backgrounds, e.g., including additional
vector or axial vector fields. In these cases the proof based on
$\ze$-regularization may be more difficult, but there are
apparently no obstacles in making explicit calculations.

\begin{acknowledgments}

One of the authors (I.Sh.) is grateful to I.L. Buchbinder for
useful discussions. The work of I.Sh. has been partially
supported by CNPq, FAPEMIG and ICTP. T.P.N. is grateful
to CAPES for supporting his Ph.D. project.
\end{acknowledgments}


\end{document}